\documentclass[structabstract]{aa}
\usepackage{graphicx}
\usepackage{txfonts}

\usepackage[authoryear,round]{natbib}

\begin{document}
   \title{Erosion of dust aggregates}


   \author{A. Seizinger,
          \inst{1}
          \and
          S. Krijt,
          \inst{2}
          \and
          W. Kley\inst{1}
          }

   \institute{Institut f\"ur Astronomie and Astrophysik, Eberhard Karls Universit\"at T\"ubingen,\\
              Auf der Morgenstelle 10, D-72076 T\"ubingen, Germany\\
              \email{alexs@tat.physik.uni-tuebingen.de}
              \and
              Leiden Observatory, Leiden University, P.O. Box 9513, 2300 RA Leiden, The Netherlands}

   \date{Received 30.09.2013; accepted 14.10.2013}

 
  \abstract
   {}
   {The aim of this work is to gain a deeper insight into how much different aggregate types are affected by erosion. Especially, it is important to study the influence of the velocity of the impacting projectiles. We also want to provide models for dust growth in protoplanetary disks with simple recipes to account for erosion effects.}
   {To study the erosion of dust aggregates we employed a molecular dynamics approach that features a detailed micro-physical model of the interaction of spherical grains. For the first time, the model has been extended by introducing a new visco-elastic damping force which requires a proper calibration. Afterwards, different sample generation methods were used to cover a wide range of aggregate types.}
   {The visco-elastic damping force introduced in this work turns out to be crucial to reproduce results obtained from laboratory experiments. After proper calibration, we find that erosion occurs for impact velocities of $5\,\mathrm{m s^{-1}}$ and above. Though fractal aggregates as formed during the first growth phase are most susceptible to erosion, we observe erosion of aggregates with rather compact surfaces as well.}
   {We find that bombarding a larger target aggregate with small projectiles results in erosion for impact velocities as low as a few $\mathrm{m s^{-1}}$. More compact aggregates suffer less from erosion. With increasing projectile size the transition from accretion to erosion is shifted to higher velocities. This allows larger bodies to grow through high velocity collisions with smaller aggregates.}


	\authorrunning{Seizinger et al.}
	\maketitle

\section{Introduction}
\label{sec:introduction}

In the past years, both laboratory experiments and numerical simulations have been able to shed light on many aspects of the growth processes leading to the formation of planetesimals. Nevertheless, various questions regarding the growth from microscopic dust grains to kilometer-sized bodies remain unanswered.

One of these open questions concerns the presence of small dust grains in protoplanetary disks. From observations, we know that sub-mm sized grains are present \citep[e.g.][]{2011ARA&A..49...67W}. Yet, theoretical growth models predict a rapid depletion of small grains by sticking \citep{2005A&A...434..971D}. Replenishing the amount of small grains during the evolution of the disk may reconcile these predictions with observations. Destructive collisions of larger bodies are likely to come to mind as a source of small particles. But other effects such as photophoresis may contribute to the production of small grains as well \citep{2006PhRvL..96m4301W, 2011ApJ...733..120K, 2013ApJ...763...11D}. Alternatively, it has been proposed that electric charging hinders the growth of larger bodies by suppressing coagulation of sub-mm sized grains \citep[e.g.][]{2009ApJ...698.1122O, 2011ApJ...731...95O}.

In this work, we perform simulations to study the erosion of different aggregates types. To model high velocity impacts more accurately, we extend the molecular dynamics approach of \citet{2012A&A...541A..59S} by a new viscoelastic damping mechanism recently presented by \citet{krijt2013}. This work is supposed to provide the necessary data for a better treatment of erosion in the existing models for dust growth in protoplanetary disks \citep[e.g.][]{2007A&A...461..215O, 2008A&A...489..931Z, 2010A&A...513A..79B, 2012A&A...540A..73W, 2013A&A...556A..37D}. Thus, we study the erosion efficiency for different aggregates with properties that are typically encountered during the growth process in protoplanetary disks.

In the beginning, dust growth is driven primarily by Brownian motion because micron-sized aggregates couple very well to the surrounding gas in the disk. Owing to the low relative velocities most frequent are hit \& stick collisions without any restructuring. Such collisions result in the growth of very fluffy, fractal aggregates \citep[e.g.][]{1996Icar..124..441B, 1999Icar..141..388K}. As these fractal aggregates are typical for the size regime of mm and below we study the erosion of fractal aggregates.

As the aggregates grow larger, their relative velocities increase and the hit \& stick regime is left. Depending on the collision velocity compaction and fragmentation will set in \citep{2000Icar..143..138B}. As of today, the further evolution of dust aggregates is hotly debated. The impact of various processes such as compaction, fragmentation, bouncing, fragmentation with mass transfer, or reaccretion in aggregate collisions at different velocities and with different porosities has been studied in numerous laboratory experiments. For a helpful summary we refer to \citet{2008ARA&A..46...21B} and \citet{2010A&A...513A..56G}.

Unfortunately, owing to the available computing power it is not possible to study any possible aggregate type at any given size. Instead, we restrict our study to a few aggregate types that may serve as prototypes. For this purpose we chose aggregates generated by particle-cluster aggregation (PCA) and several ballistic-aggregation-and-migration (BAM) aggregates. The results obtained for these aggregate types are supposed to give an estimate on the erosion efficiencies expected for the more compact aggregates formed during the growth process in a protoplanetary disk.

\section{Interaction model}

\subsection{Established model}
\label{sec:interaction_model}

In our simulations, aggregates are composed of thousands of equal sized, spherical grains (also referred to as monomers). Monomers interact with each other only if they are in contact. Energy is dissipated upon deformation of these contacts caused by the relative motion of the grains. Long range forces such as electromagnetic forces or gravity are not taken into account.

\begin{figure}
\resizebox{\hsize}{!}{\includegraphics{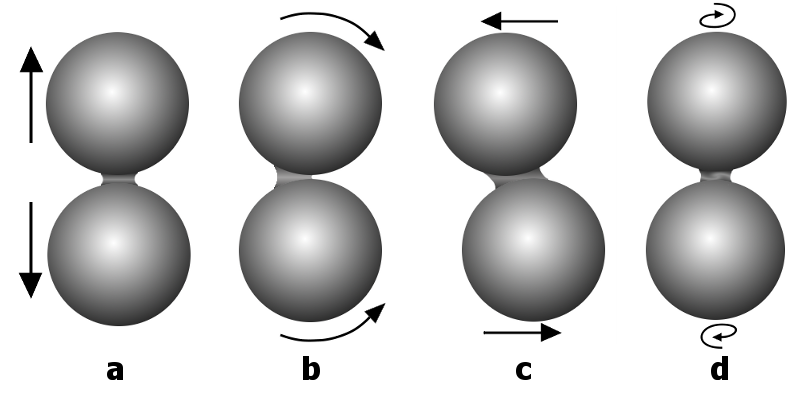}}
\caption{The four types of particle interaction: Compression/Adhesion (a), Rolling (b), Sliding (c), and Twisting (d). Figure taken from \citet{2012A&A...541A..59S}.}
\label{fig:interaction_model}
\end{figure}

We use nearly the same interaction model as proposed by \citet{1997ApJ...480..647D}. To model the interaction of two spherical grains they distinguish between four types of motions (see Fig.\ref{fig:interaction_model}). The equations describing these types of motions are mostly based on earlier theoretical work \citep{1971RSPSA.324..301J,1995PMagA..72..783D, 1996PMagA..73.1279D}. For rolling, sliding, and twisting, the interaction remains elastic as long as the displacement from the equilibrium state remains small. If a certain threshold is exceeded, the motion enters the inelastic regime and energy is being dissipated. Apart from one minor difference \citet{2007ApJ...661..320W} derived the same equations from corresponding potentials. This brings the advantage of being able to track how much energy is dissipated by which type of motion. 

However, compared to laboratory experiments on the compression of porous dust aggregates performed by \citet{2009ApJ...701..130G} the behavior predicted by the model of \citet{1997ApJ...480..647D} was too soft. To overcome this discrepancy, \citet{2012A&A...541A..59S} modified the rolling and sliding interaction. They observed much better agreement between simulations and laboratory results by increasing the rolling interaction by a factor of $8$ and the sliding interaction by a factor of $2.5$.

In this work, we employ the modified interaction model proposed by \citet{2012A&A...541A..59S} with $m_{\mathrm{r}} = 8$ and $m_{\mathrm{s}} = 2.5$. The material parameters are listed in Tab.\,\ref{tab:material_parameters}.

\begin{table}
 \caption[]{Material parameters.}
 \label{tab:material_parameters} 
 \centering
 \renewcommand\arraystretch{1.2}
 \begin{tabular}{ll}
   \hline
   \noalign{\smallskip}
   Physical property & Silicate\\
   \noalign{\smallskip}
   \hline
   \noalign{\smallskip}
   Particle Radius $r$ (in $\mathrm{\mu m}$) & $0.6$\\
   Density $\rho$ (in g\, cm$^{-3}$) & $2.65$\\
   Surface Energy $\gamma$ (in mJ\, m$^{-2}$) & $20$\\
   Young's Modulus $E$ (in GPa) & $54$\\
   Poisson Number $\nu$ & $0.17$\\
   Critical Rolling Length $\xi_\mathrm{crit}$ (in nm) & $2$\\
   Viscous damping time $T_{\mathrm{vis}}$ (in s) & $1.25 \cdot 10^{-11}$\\
   \noalign{\smallskip}
   \hline
 \end{tabular}
\end{table}

\subsection{Visco-elastic damping}
The critical sticking velocity $v_\mathrm{crit}$ at which the transition from sticking to bouncing occurs constitutes an important value when comparing the collisional behavior predicted by a theoretical interaction model with laboratory results. For micron sized silicate grains JKR theory predicts $v_\mathrm{crit} \approx 0.1\,\mathrm{m s^{-1}}$. However, in laboratory experiments on the stickiness of such grains a considerably higher sticking velocity of the order of $1\,\mathrm{m s^{-1}}$ has been measured \citep{2000ApJ...533..454P}.

As an attempt to overcome this discrepancy \citet{2008A&A...484..859P} proposed surface asperities as a possible damping mechanism. Upon collision of two monomers small asperities on their surfaces get flattened. The corresponding plastic deformation would lead to the additional dissipation of kinetic energy. The damping was applied by artificially lowering the relative velocity of two monomers in the integration step where they collided with each other. However, when performing simulations with higher collisions velocities ($> \mathrm{ms^{-1}}$) \citet{2012A&A...541A..59S} found that this damping mechanism introduced numerical instability.

In this work, we instead use the new damping force derived by \citet{krijt2013}, who show that for viscoelastic materials, the dissipative stresses in the contact area can be integrated to yield a damping force
\begin{equation}F_\mathrm{D} = \frac{2 T_\mathrm{vis} E^\star}{\nu^2} a\,v_\mathrm{rel},\label{eq:damping_force}\end{equation}
where $a$ denotes the current contact radius and $v_\mathrm{rel}$ the relative normal velocity of the two monomers. The Poisson number $\nu$ and the reduced Young's modulus $E^\star = E / (2 (1 - \nu^2))$ are material constants. The viscoelastic timescale $T_\mathrm{vis}$ is not well-known, but values around $10^{-12} \-- 10^{-11}\,\mathrm{s}$ allowed \citet{krijt2013} to reproduce collision experiments with single microspheres very well.

The damping force given in Eq.\,\ref{eq:damping_force} replaces the weak damping introduced by \citet{2012A&A...541A..59S} to prevent aggregates from being heated up artificially \citep{2008A&A...484..859P}.

\section{Erosion of RBD cakes}
\subsection{Calibration}
\label{sec:calibration}

In the first step, we calibrate our extended interaction model using the results of laboratory experiments performed by \citep{2011ApJ...734..108S}. In their work, they shot a volley of single monomers on a sample dust cake (from now on referred to as projectiles and target). The samples have been generated by random ballistic deposition (RBD) and had a high porosity \citep{PhysRevLett.93.115503}. The velocity of the incoming projectiles was $15$, $30$, $45$, and $60\,\mathrm{m s^{-1}}$. After shooting a certain number of projectiles at the target the current weight of the target was measured. By repeating this procedure they determined the evolution of the mass loss with respect to the total projectile mass exposure \citep[see][Fig.\,4]{2011ApJ...734..108S}.

In our simulations, we try to follow this procedure as closely as possible. We start by generating a target via random ballistic deposition. Owing to the computational demand imposed by high numbers of particles we limit the base area of the target to $100 \times 100\,\mathrm{\mu m}$. Initially, the target has a height of $\approx 70\,\mathrm{\mu m}$ and is composed of $10^5$ monomers. Then, a barrage of 100 randomly distributed monomers is shot at the target. The mass loss $\Delta m$ of the target is given by the number of monomers $\Delta N$ that are knocked out by the incoming projectiles. Dividing the number of eroded monomers by the number of projectiles we obtain the erosion efficiency $\epsilon$
\begin{equation}\epsilon = \frac{\Delta N}{N_\mathrm{p}} ,\end{equation}
where $N_\mathrm{p}$ denotes the total number of monomers of the incoming projectiles and $\Delta N = N_\mathrm{target, before} - N_\mathrm{target, after}$ the change of the number of monomers of the target aggregate. Thus, for $\epsilon > 0$ the target has been eroded whereas for $\epsilon < 0$ some of the projectile mass has been accreted onto the target. The values can be compared directly with the results from \citet{2011ApJ...734..108S}. 

\begin{figure}
\resizebox{\hsize}{!}{\includegraphics{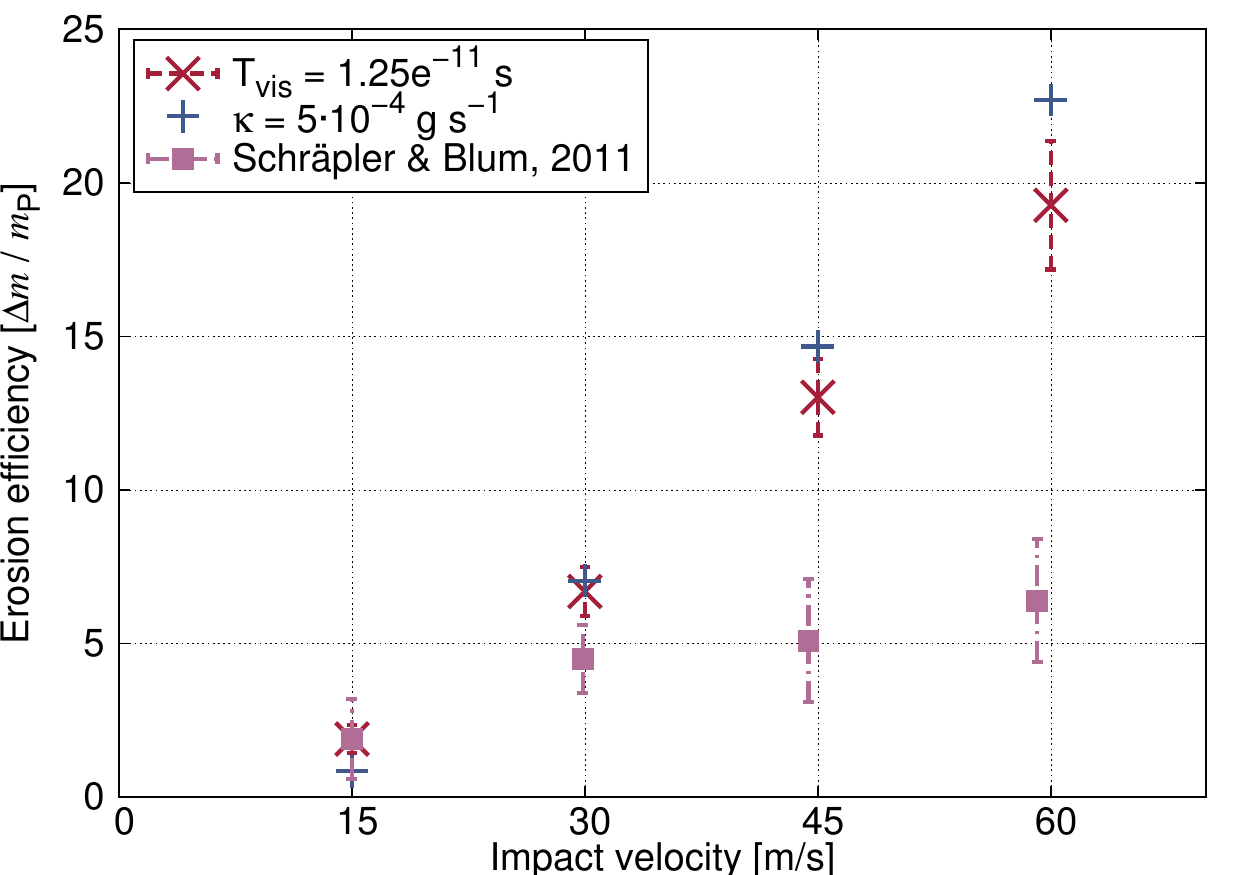}}
\caption{Erosion efficiency for different velocities of the incoming projectiles. The errorbars have been determined by using 6 randomly generated targets with equal properties. At velocities below $30\,\mathrm{m s^{-1}}$ the results we obtain when setting $T_\mathrm{vis} = 1.25\cdot10^{-11}\,\mathrm{s}$ (red crosses) agree well with laboratory experiments by \citet{2011ApJ...734..108S} (purple squares). At higher velocities our micromechanical interaction model is probably no longer applicable as plastic deformation of the monomers will play a larger role. For comparison, we show results obtained by increasing the strength of normal damping mechanism introduced in \citet{2012A&A...541A..59S} by a factor of 500 (blue crosses). }
\label{fig:cake_erosion_efficiency}
\end{figure}

In Fig.\,\ref{fig:cake_erosion_efficiency}, the erosion efficiency obtained from our simulations is compared to laboratory results \citep[][Fig.\,5]{2011ApJ...734..108S}. By choosing $T_\mathrm{vis} = 1.25\cdot10^{-11}\,\mathrm{s}$ we get an erosion efficiency of $1.89 \pm 0.45$ for $v = 15\,\mathrm{m s^{-1}}$. This is in excellent agreement with the value of $1.9 \pm 1.3$ obtained from laboratory experiments \citep{2011ApJ...734..108S}. If we perform the simulations without the additional visco-elastic damping force we find $\epsilon = 83.16 \pm 1.68$ for $v = 15\,\mathrm{m s^{-1}}$. This demonstrates impressively why a proper treatment of such damping effects is crucial in the high velocity regime. 

Finding a value for $T_\mathrm{vis}$ that fitted the whole velocity range well was not possible. We believe that the discrepancy at collision velocities $\approx 30\,\mathrm{m s^{-1}}$ and above is caused by the plastic deformation of single monomers. A plastic yield velocity of $30\,\mathrm{m s^{-1}}$ implies a material yield strength of $\approx 3\,\mathrm{GPa}$ \citep{Thornton1998154}, which is well within the range of $0.1 \-- 11\,\mathrm{GPa}$ given by \citet{2008A&A...484..859P}. In this velocity regime our physical model may therefore not provide a good description anymore. Thus, we put our focus on the data points for lower impact velocities, which are also more relevant in the context of the collisions of smaller aggregates in protoplanetary disks \citep[e.g.][]{2008A&A...480..859B}.

For comparison, we also show results from simulations without the new damping force given in Eq.\,\ref{eq:damping_force}. Instead, we greatly increased the strength of the normal damping force by setting $\kappa = 5 \cdot 10^{-4}\,\mathrm{g s^{-1}}$ \citep[see][Sect.\,2.1.4]{2012A&A...541A..59S}. This corresponds to an increase of $\kappa$ by a factor of $500$. For $v = 15\,\mathrm{m s^{-1}}$ this results in a drop of the erosion efficiency drops from $83.16$ to $0.85$. This means that we can also get much closer to the laboratory results by greatly increasing the strength of the normal damping (see blue crosses in Fig.\,\ref{fig:cake_erosion_efficiency}). Nevertheless, we prefer the viscoelastic damping force proposed by \citet{krijt2013} because its derivation is based on physical deliberations, whereas the weak normal damping had been introduced for numerical reasons only.

\subsection{Passivation}
\label{sec:passivation}

\begin{figure}
\resizebox{\hsize}{!}{\includegraphics{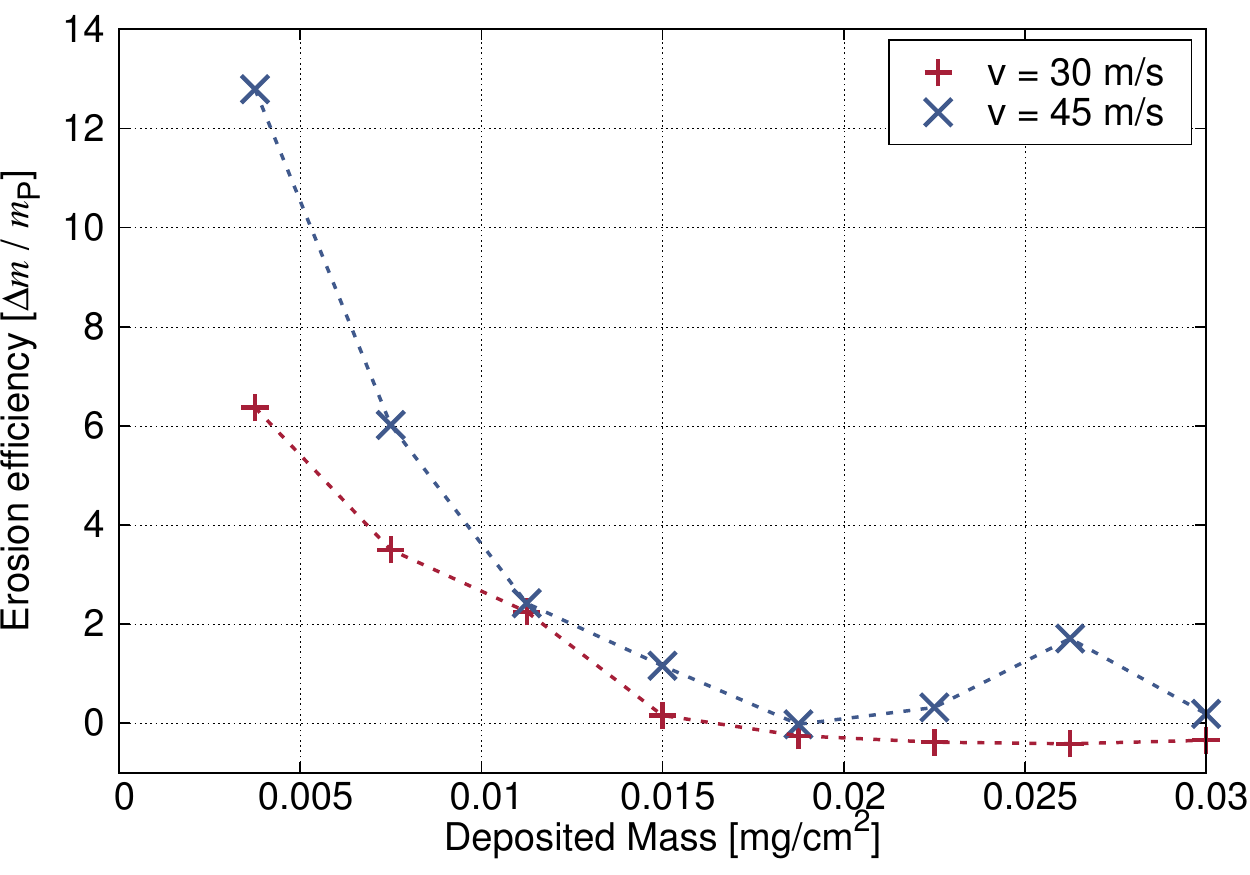}}
\caption{Erosion efficiency with respect to the cumulative mass of the projectiles that have been shot at the target. Like in the laboratory experiments the trajectory of the incoming projectiles is perpendicular to the surface of the target. The target aggregate becomes passivated quickly which leads to a significant drop of the erosion efficiency.}
\label{fig:cake_passivation_plot}
\end{figure}

\begin{figure*}
\resizebox{\hsize}{!}{\includegraphics{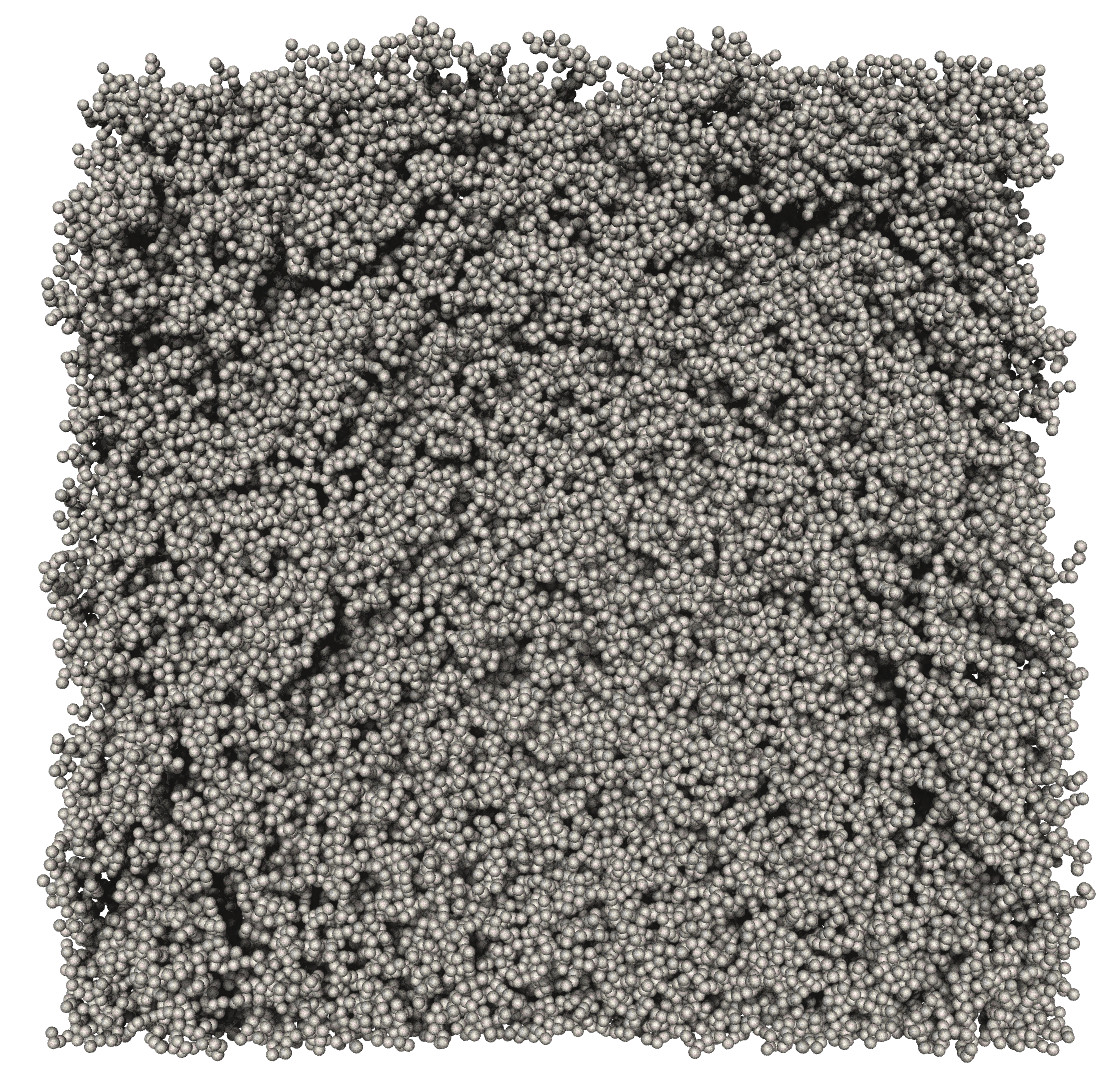} \hfill \includegraphics{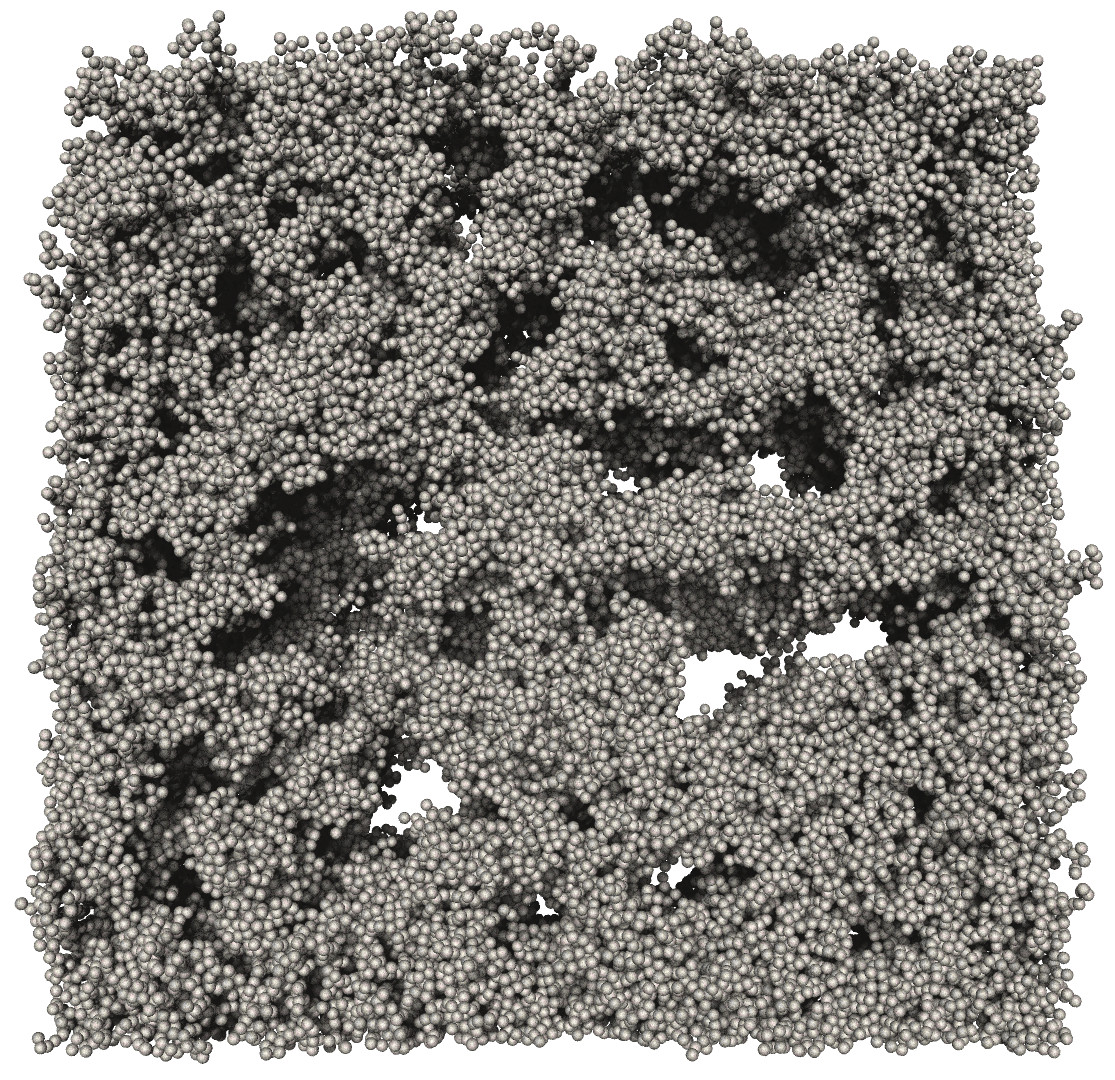}}
\caption{Top-down view on a target aggregate before and after an impact sequence. Following the bombardment with a sufficient number of high speed projectiles the shape of the target aggregate changes significantly. \textit{Left:} Initially, the RBD sample features a very homogeneous structure. \textit{Right:} After shooting $1000$ projectiles at the sample with $v = 30\,\mathrm{m s^{-1}}$ the homogeneous structure has been destroyed. Deep holes and more compact pillars have formed.}
\label{fig:cake_passivation_image}
\end{figure*}

The experiments by \citet{2011ApJ...734..108S} show a decline of the erosion efficiency to very low values after shooting in a sufficiently large number of projectiles. Successive impacts restructure the upper layers of the target in such a way that further projectiles are less likely to knock out monomers. To check if we can reproduce this effect we bombard the same target repeatedly with 100 monomers. For each barrage we measure the erosion efficiency independently. Indeed, we observe a similar passivation effect (see Fig.\,\ref{fig:cake_passivation_plot}). An example of how the structure of the samples changes after a bombardment with $1000$ projectiles at $v = 30\,\mathrm{m s^{-1}}$ is depicted in Fig.\,\ref{fig:cake_passivation_image}. As shown in the right panel of Fig.\,\ref{fig:cake_passivation_image} the bombardment leads to the formation of deep holes and pillar like structures. Compared to the initial density in a RBD aggregate the monomer density in these pillars is increased. Similar restructuring processes have also been observed in the laboratory experiments \citep[see][Fig.\,6]{2011ApJ...734..108S}.

So far, the trajectories of all projectiles have been perpendicular to the surface of the target. This leads to the question whether the pillar-shaped features will also emerge when projectiles impact under different angles. While shooting in monomers from random directions is complicated in laboratory experiments it does not pose any problem in numerical simulations. Apart from the impact direction we use the same setup as before.

\begin{figure}
\resizebox{\hsize}{!}{\includegraphics{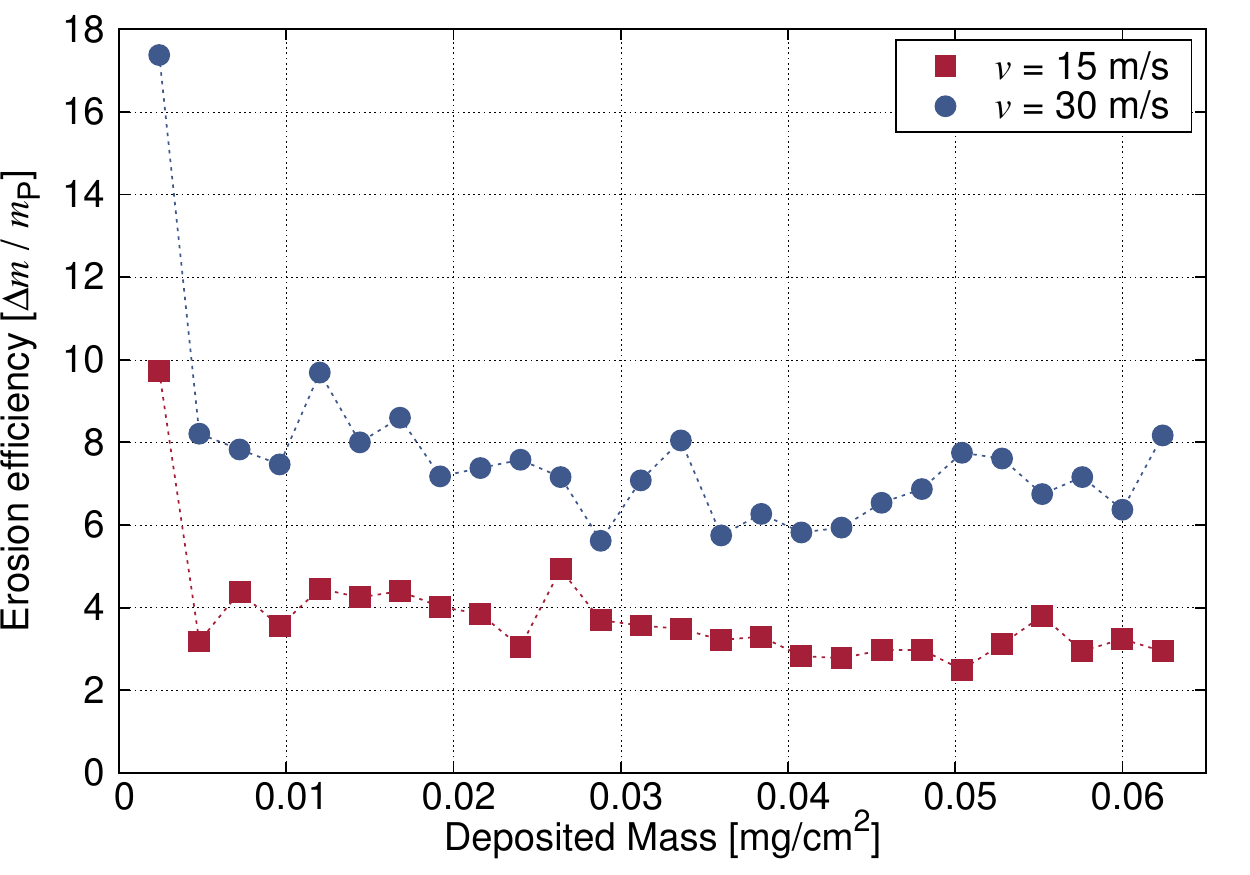}}
\caption{Erosion efficiency with respect to the cumulative mass of the projectiles. Because the projectiles are coming in from random directions, this time a larger cake with a base area of $150 \times 150\,\mathrm{\mu m}$ and a height of $\approx 70\,\mathrm{\mu m}$ is used. Each data point depicts the average erosion efficiency after shooting in a bunch of 100 monomers. The random distribution of impact angles greatly reduces the passivation effect. At the end of the simulations roughly a quarter of the aggregate mass has been eroded.}
\label{fig:erosion_passivation_rand}
\end{figure}

Indeed, after randomizing the impact angles the passivation effect seems to vanish (see Fig.\,\ref{fig:erosion_passivation_rand}). One might argue that there is small decline of the erosion efficiency, which indicates that the simulations had not fully converged yet. However, at the end of the simulations $25\,\%$ of the initial target mass have already been eroded. As shown later in Sect.\,\ref{sec:results}, no passivation is observed when shooting projectiles at an aggregate from random directions. Therefore we may conclude that it originates from the specific setup of the laboratory experiments. This is a relevant result because in the context of planet formation projectiles will hit from random directions. Thus, at least on the microscopic scale, passivation does not play an important role.

However, in Fig.\,\ref{fig:erosion_passivation_rand} we notice a strong decrease of the erosion efficiency between the first and the second barrage of projectiles. Presumably, this is caused by chopping off the uppermost fractal chains of the initial target cake. As a results of the RBD generation process the uppermost part of the aggregate is less homogeneous than the parts below. Fractal, very fluffy chains of monomers stick out. They can be sandblasted away very easily by tangential hits. The first barrage of projectiles is sufficient to erode this upper layer.

Compared to the calibration simulations (see Fig.\,\ref{fig:cake_erosion_efficiency}) we obtain higher values for the erosion efficiency when the projectiles impact from random directions. For $v = 15\,\mathrm{m s^{-1}}$ the erosion efficiency roughly doubles from $\epsilon_{\mathrm{calibration}} = 1.9$ to $\epsilon_{\mathrm{random}} \approx 4$. The rise of the erosion efficiency is no surprise: In the calibration setup projectiles hit the surface of the target under an angle of $90^{\circ}$. Their kinetic energy suffices to knock a few monomers out at the impact location. The majority of these monomers is pushed deeper into the sample where they may be recaptured because their excess kinetic energy is dissipated by subsequent collisions with other monomers of the target. Since this recapturing mechanism is less effective for tangential impacts we measure a higher erosion efficiency when shooting in projectiles from random directions.

\section{Erosion of aggregates}

In this section we extend our studies to a variety of more realistic aggregates. The setup for the calibration simulation is somewhat artificial compared to the processes in a protoplanetary disk. However, bombarding a free floating aggregate with single projectiles at a well defined impact velocity in laboratory experiments is not possible at the present time. Thus, we employ numerical simulation to address this question.

\subsection{Sample generation}
\label{sec:sample_generation}

\begin{figure*}
\resizebox{\hsize}{!}{\includegraphics{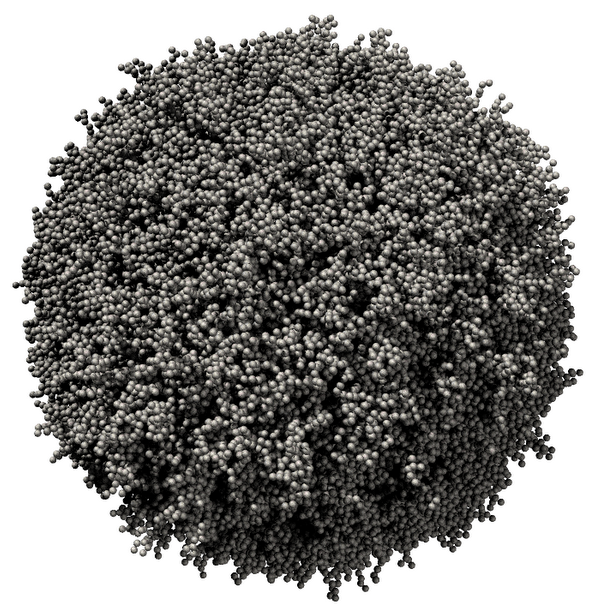} \hfill \includegraphics{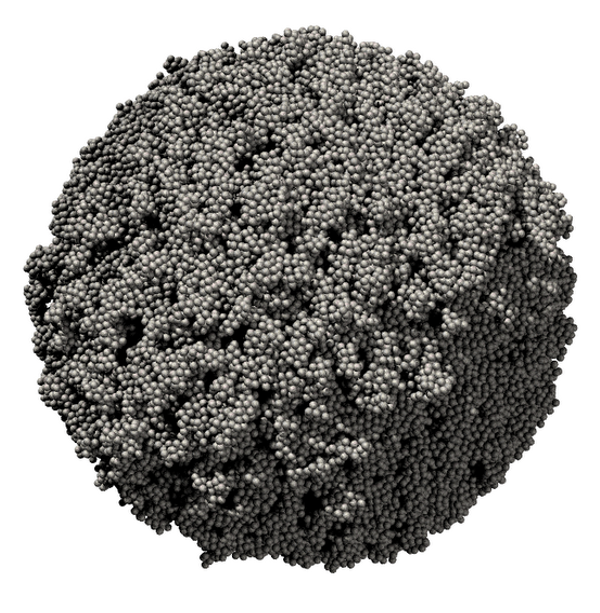}  \hfill \includegraphics{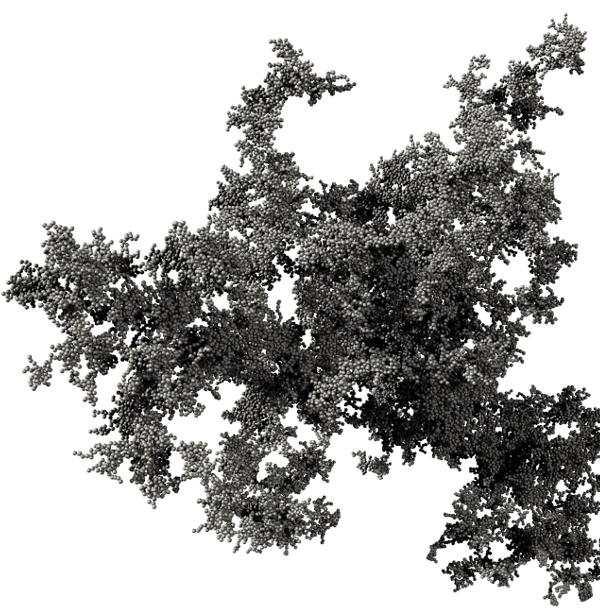}}
\caption{Sample aggregates used in this work. \textit{Left:} PCA aggregate ($n_\mathrm{c} = 2$) with $8 \cdot 10^{4}$ monomers and a diameter of $100\,\mathrm{\mu m}$. \textit{Center:} BAM aggregate consisting of $1.5 \cdot 10^{5}$ monomers with $n_\mathrm{c} = 6$ and a diameter of $100\,\mathrm{\mu m}$. \textit{Right:} Fractal aggregate ($n_\mathrm{c} \approx 2$) consisting of $6 \cdot 10^{4}$ monomers and a maximum diameter of $\approx 280\,\mathrm{\mu m}$.}
\label{fig:aggregates}
\end{figure*}

The target aggregates used in this work have been generated by a variety of methods. The aggregates generated by particle-cluster aggregation and random ballistic deposition are spherical and homogeneous. In contrast, the fractal aggregates have a non-spherical shape and are highly inhomogeneous. Examples of all three types are shown in Fig.\,\ref{fig:aggregates}.

Particle-cluster aggregation (PCA) constitutes an easy way to generate larger aggregates. The aggregate grows by adding single monomers from random directions. The monomers stick at the location where the first contact with the existing aggregate is established. Thus, the resulting aggregate will be rather fluffy with a volume filling factor of $\phi \approx 0.19$ (see left panel of Fig.\,\ref{fig:aggregates}). This procedure is similar to random ballistic deposition except that particles are coming from random directions rather than a specific side.

The second aggregate type is ballistic aggregation and migration (BAM) which has been suggested by \citet{2008ApJ...689..260S}. As in the case of PCA monomers approaching from random directions are successively added to the aggregate. However, the final position of a monomer is determined in such a way that contact to two or three existing monomers is established at the same time resulting in more compact aggregates. For a more detailed description of the generation process we refer to \citet[][Sect.\,3.2]{2013A&A...551A..65S}. In this work, we use two-times migration (BAM2), which means that after migrating once to establish contact with a second monomer, the migration process is repeated to get in contact with a third monomer. This procedure generates compact aggregates with a coordination number $n_\mathrm{c} = 6$. An example of such an aggregate is shown in the center of Fig.\,\ref{fig:aggregates}.

The two aggregate types described above both share the disadvantage that their structure is somewhat artificial. Thus, we also use aggregates which have been obtained from a joint project where two different numerical techniques have been combined to simulate the growth of dust aggregates (Seizinger et al., in prep.). Starting with aggregates consisting of a single monomer, we followed the evolution of a swarm of representative aggregates using the approach presented by \citet{2008A&A...489..931Z}. On the microscopic scale, every collision between two representative aggregates has been simulated using molecular dynamics. That way, the changes of the aggregate structure during the growth process could be resolved in great detail. The growth of sub-mm sized aggregates is primarily driven by Brownian motion which results in very porous aggregates \citep[e.g.][]{1999Icar..141..388K}. An example is depicted in the right panel of Fig.\,\ref{fig:aggregates}.

As already mentioned in Sect.\ref{sec:introduction} these aggregates have been chosen as prototypes reflecting different stages in the evolution of dust aggregates. When dealing with dust aggregates of different porosities/structure the following equations may serve as easily implementable recipes to account for erosion.

\subsection{Results}
\label{sec:results}

Using the sample aggregates described in Sect.\,\ref{sec:sample_generation} we determine the erosion efficiency in the following way:

First, a given number of single monomers (from now on referred to as projectiles) is randomly distributed around the target in such a way, that their trajectories will hit the target with an impact parameter $b$ between 0 and 1. The impact parameter is chosen such that the number of impacts per cross section area are constant. To avoid projectiles interfering with each other we restrict the total number of incoming projectiles to $20$. A lower number of projectiles is used when increasing their size in Sect.\,\ref{sec:projectile_size}. The erosion efficiency is calculated in the same way as described in Sect.\,\ref{sec:calibration}.

We perform simulations for impact velocities from $1\,\mathrm{m s^{-1}}$ to $15\,\mathrm{m s^{-1}}$. Note that the velocity range has been chosen to compare our results to the calibration experiments. Even in turbulent disks impact velocities of $15\,\mathrm{m s^{-1}}$ are quite high for mm-sized aggregates \citep[e.g.][]{2008A&A...480..859B}.

For each velocity, we perform 5 simulations with a different initial distribution of the projectiles and average over the results. As the fractal aggregates have a inhomogeneous density, we use 5 different aggregates of similar size / structure.

\begin{figure}
\resizebox{\hsize}{!}{\includegraphics{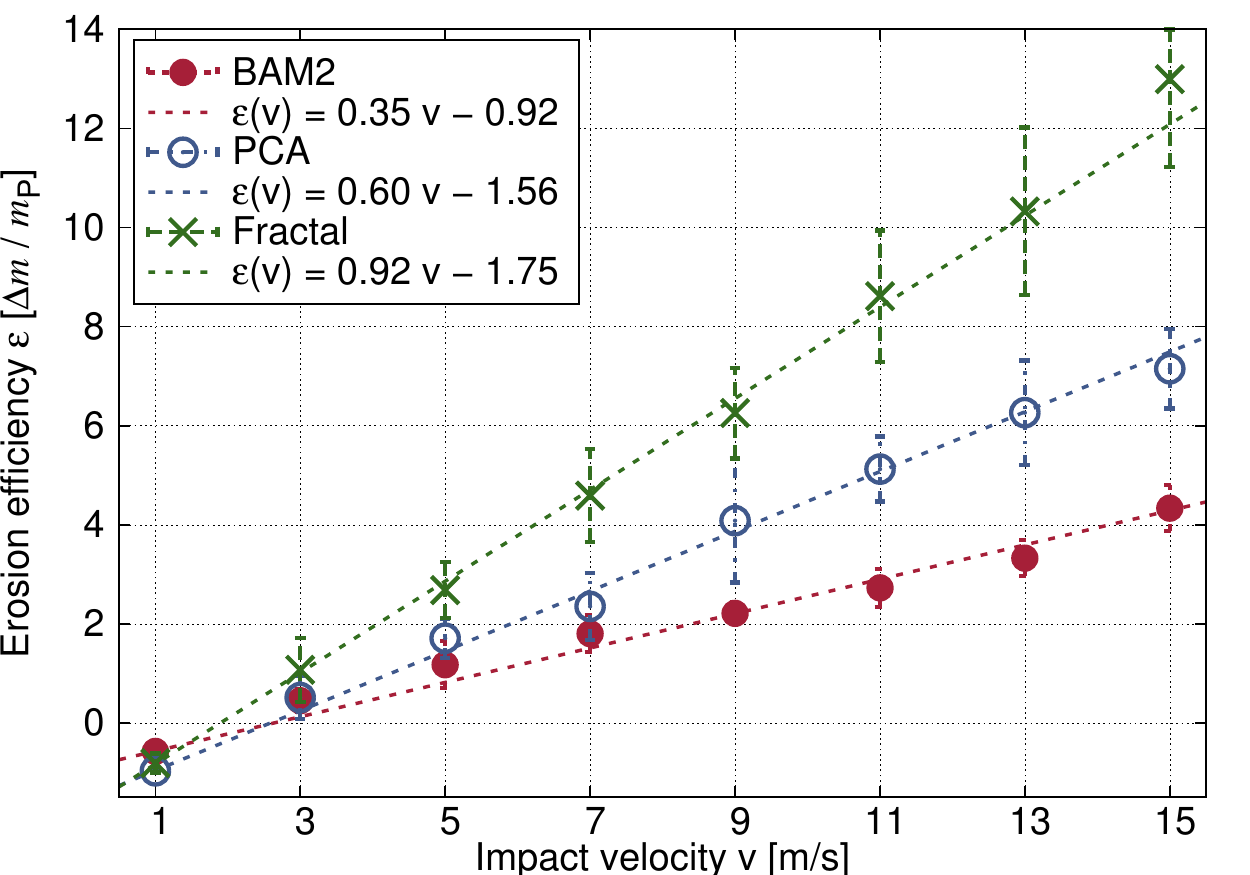}}
\caption{Erosion efficiency for different types of aggregates. In general, the erosion efficiency is lower for compact aggregates. The threshold velocity where two monomers stick to each other is $2.6\,\mathrm{m s^{-1}}$. Around this velocity we observe the transition from accretion to erosion.}
\label{fig:agg_erosion_efficiency}
\end{figure}

The results are shown in Fig.\,\ref{fig:agg_erosion_efficiency}. For low velocities the erosion efficiency approaches a value of $-1$ which corresponds to accretion rather than erosion. For both, the compact BAM2 and the rather porous PCA aggregate, the transition from accretion to erosion occurs at an impact velocity of $v \approx 2\,\mathrm{m s^{-1}}$. As one would expect we observe a lower erosion efficiency for the ``hardened'' BAM2 aggregates. For $v = 15\,\mathrm{m s^{-1}}$  the erosion efficiency measured for any of the target aggregates is well above the corresponding value of $1.9$ obtained from the calibration simulations. As already explained in the last paragraph of Sect.\,\ref{sec:passivation}, this is expected when the target is bombarded from random directions.

To derive simple recipes for the dependency of the erosion efficiency $\epsilon$ on the impact velocity $v$ we determined fit curves for the different aggregate types. Based on the results shown in Fig.\,\ref{fig:agg_erosion_efficiency} we chose a linear fit. We find
\begin{eqnarray}\epsilon_\mathrm{BAM2}(v) = 0.35 v - 0.92,\end{eqnarray}
\begin{eqnarray}\epsilon_\mathrm{PCA}(v) = 0.60 v - 1.56,\end{eqnarray}
\begin{eqnarray}\epsilon_\mathrm{frac}(v) = 0.92 v - 1.75,\end{eqnarray}
where $v$ is given in $\mathrm{m s^{-1}}$. Note that these fits should be applied with care for velocities below $1\,\mathrm{m s^{-1}}$. Negative values of $\epsilon(v)$ correspond to accretion, where $\epsilon = -1$ means that all incoming projectiles have been accreted onto the target. Obviously, values below $-1$ do not represent any physical process and are just an artifact of the fitting process. As accretion is dominating for low velocities, it applies 
\begin{eqnarray}\lim\limits_{v \rightarrow 0} \epsilon(v) = -1.\end{eqnarray}

It is important to note that the erosion efficiency for the fractal aggregates has been determined in a different way than explained in Sect.\,\ref{sec:calibration}. As the impact velocities of the projectiles increase, sometimes whole ``fractal arms'' are chopped off the main aggregate. However, this process resembles fragmentation rather than erosion. Thus, we only count the monomers of fragments that consist of fewer than 10 monomers when determining the erosion efficiency. In case of the more compact aggregates fragments consisting of more than two to three monomers are very rarely to be found. Thus, the both methods to determine the erosion efficiency return the same values.

\begin{figure}
\resizebox{\hsize}{!}{\includegraphics{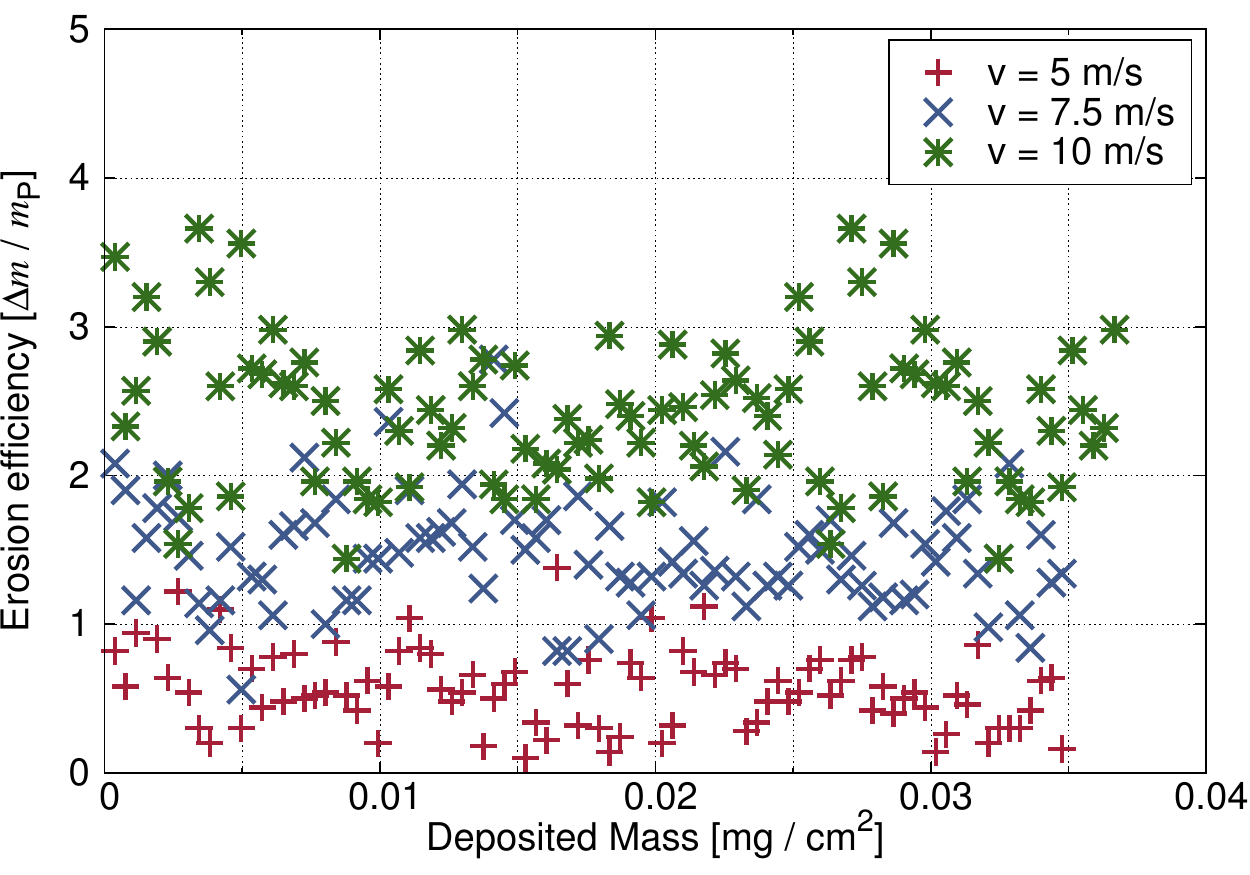}}
\caption{Evolution of the erosion efficiency with an increasing amount of the deposited mass. The target was a PCA aggregate with a diameter of $100\,\mathrm{\mu m}$. Each data point corresponds to the average erosion efficiency measured after shooting in 100 monomers. Contrary to the calibration setup no passivation effect is observed.}
\label{fig:agg_passivation}
\end{figure}

We were curious whether we could reproduce the passivation effect we observed in our calibration simulations for aggregates. For this purpose we repeatedly shot 50 projectiles at the PCA aggregate (left panel of Fig.\,\ref{fig:aggregates}). After every barrage the erosion efficiency was measured and the fragments were removed. We choose the PCA aggregate because its structure is very similar to the RBD samples used for the calibration. However, no decline of the erosion efficiency was observed (see Fig.\,\ref{fig:agg_passivation}). Initially, the average coordination number of the PCA aggregate is 2. As a result of the bombardment monomers are pushed inwards which leads to an increase of the average coordination number of the aggregate. This supports the assumption that the ``holes and pillars'' shown in Fig.\,\ref{fig:cake_passivation_image} play a key role for the accretion of incoming projectiles.

\subsection{Influence of the projectile size}
\label{sec:projectile_size}

So far, the projectiles consisted of only a single monomer. However, aggregates of different sizes will be present in a protoplanetary disk. Thus, we extend our study to larger projectiles which are generated by particle-cluster aggregation. The size of the projectile aggregates lies between two and a few thousand monomers, which means that their mass remains at least an order of magnitude below the target mass.

\begin{figure}
\resizebox{\hsize}{!}{\includegraphics{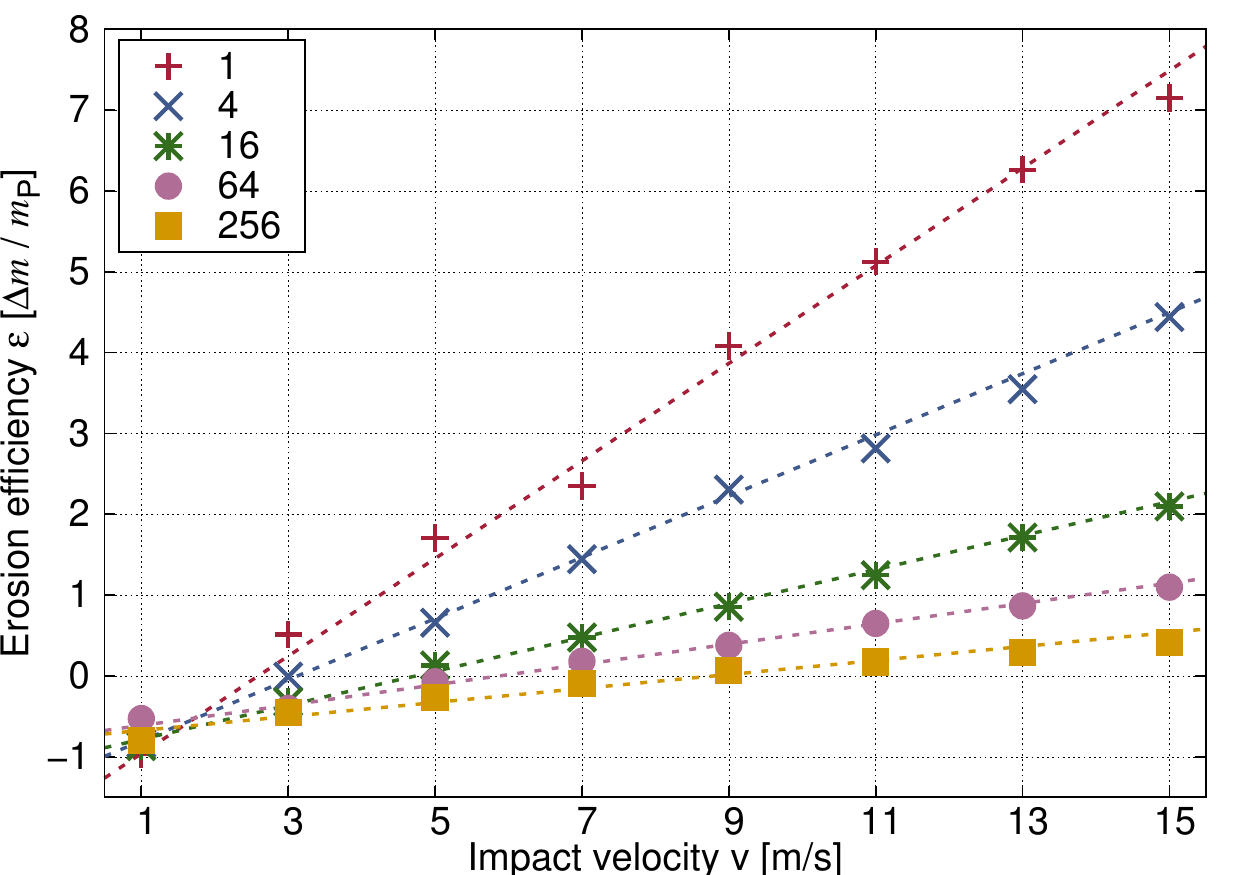}}
\caption{Comparison of the erosion efficiency for different projectile sizes (in monomers). The PCA target has the same properties as the one depicted in Fig.\,\ref{fig:aggregates}. With growing projectile mass the transition from accretion to erosion is shifted to higher velocities.}
\label{fig:projectile_size}
\end{figure}

\begin{figure*}
\resizebox{\hsize}{!}{\includegraphics{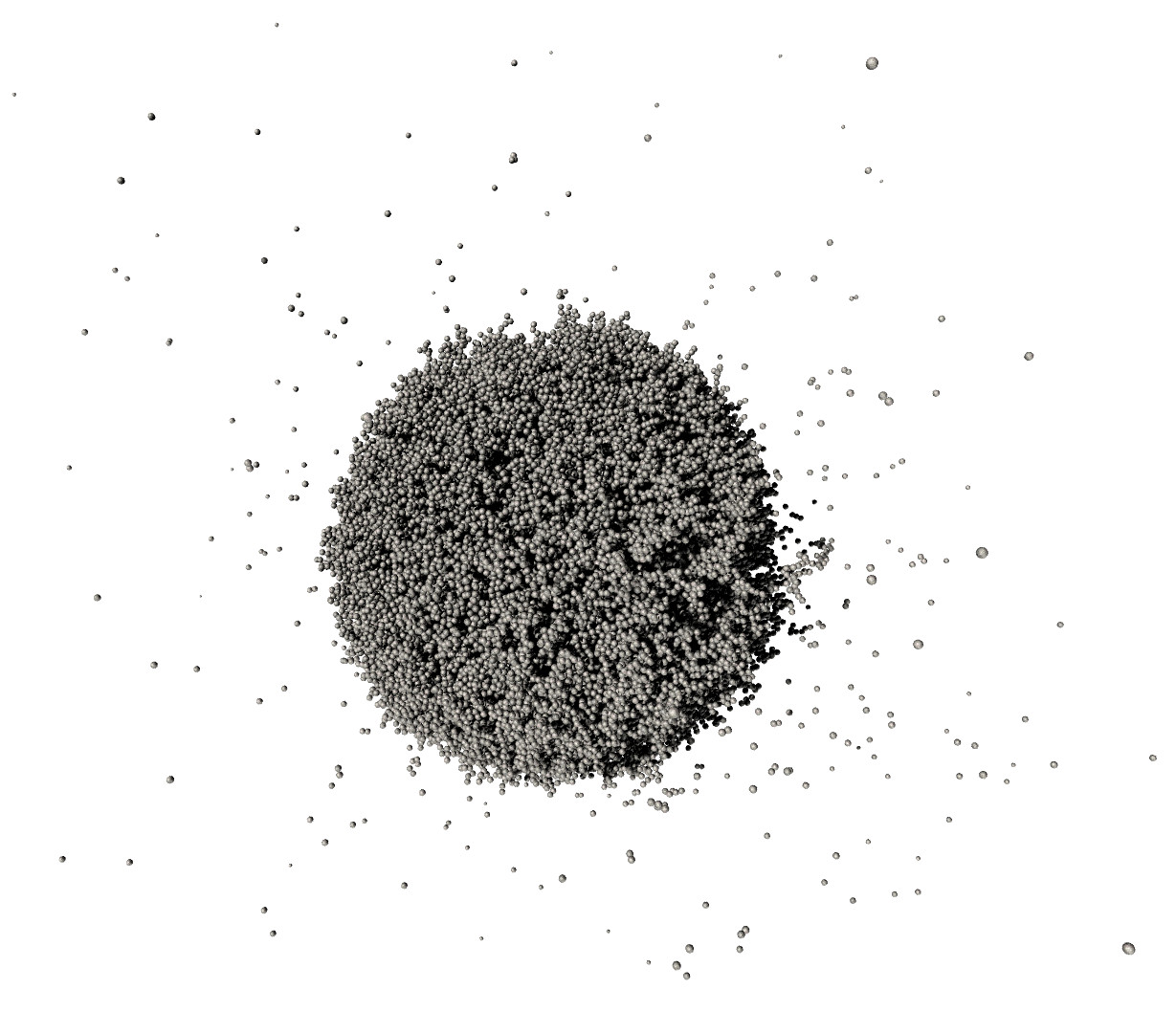} \hfill \includegraphics{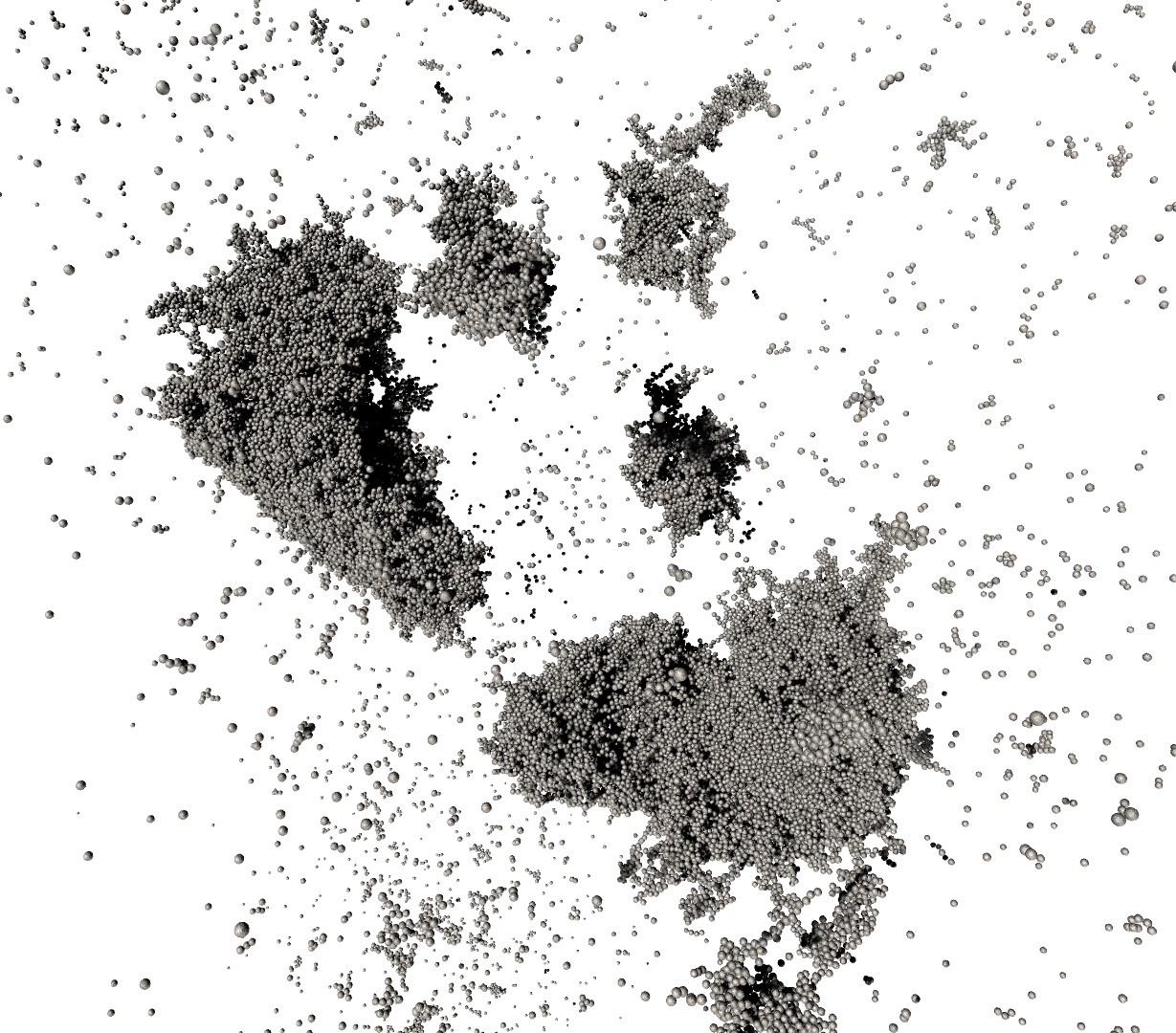}}
\caption{Depending on the size of the projectiles the outcome changes from erosion to fragmentation. In both cases the impact velocity was $15\,\mathrm{m s^{-1}}$. \textit{Left:} Erosion after shooting in 25 projectiles each consisting of 4 monomers. \textit{Right:} Fragmentation after shooting in 4 projectiles each consisting of 4096.}
\label{fig:erosion_vs_fragmentation}
\end{figure*}

In the following simulations the PCA aggregates with the properties specified in Fig.\,\ref{fig:aggregates} serve as targets. As shown in Fig.\,\ref{fig:projectile_size} the projectile size heavily influences the outcome of our simulations. For increasing size of the projectiles the erosion efficiency drops. At first glance, this may seem counterintuitive. The key to understand this observation lies in the, compared to the target, low mass of projectile. While the kinetic energy of a single monomer suffices to knock a few monomers out of the target aggregate, it is vastly below the energy threshold required to disrupt the entire aggregate. When the first monomers of the projectile hit the target monomers are knocked out. However, subsequent parts of the incoming projectile may push these eroded monomers back toward the target where they may get reaccreted. For small projectiles there is a higher probability for eroded monomers to escape from the target aggregate.

In the context of planet formation the growth of larger bodies is a key issue. Therefore we are especially interested in the velocity $v_{\mathrm{A}\rightarrow\mathrm{E}}$ at which the transition from accretion to erosion occurs. For this purpose we first determine linear fits $\epsilon(v) = a v + b$ for the results as depicted in Fig.\,\ref{fig:projectile_size}. Then, $v_{\mathrm{A}\rightarrow\mathrm{E}}$ can be calculated via $v_{\mathrm{A}\rightarrow\mathrm{E}} = -(b/a)$. To examine the influence of the porosity of the target, we used rather fluffy PCA aggregates (see Fig.\,\ref{fig:projectile_size}) as well as compact BAM2 aggregates (not shown) as targets.

\begin{figure}
\resizebox{\hsize}{!}{\includegraphics{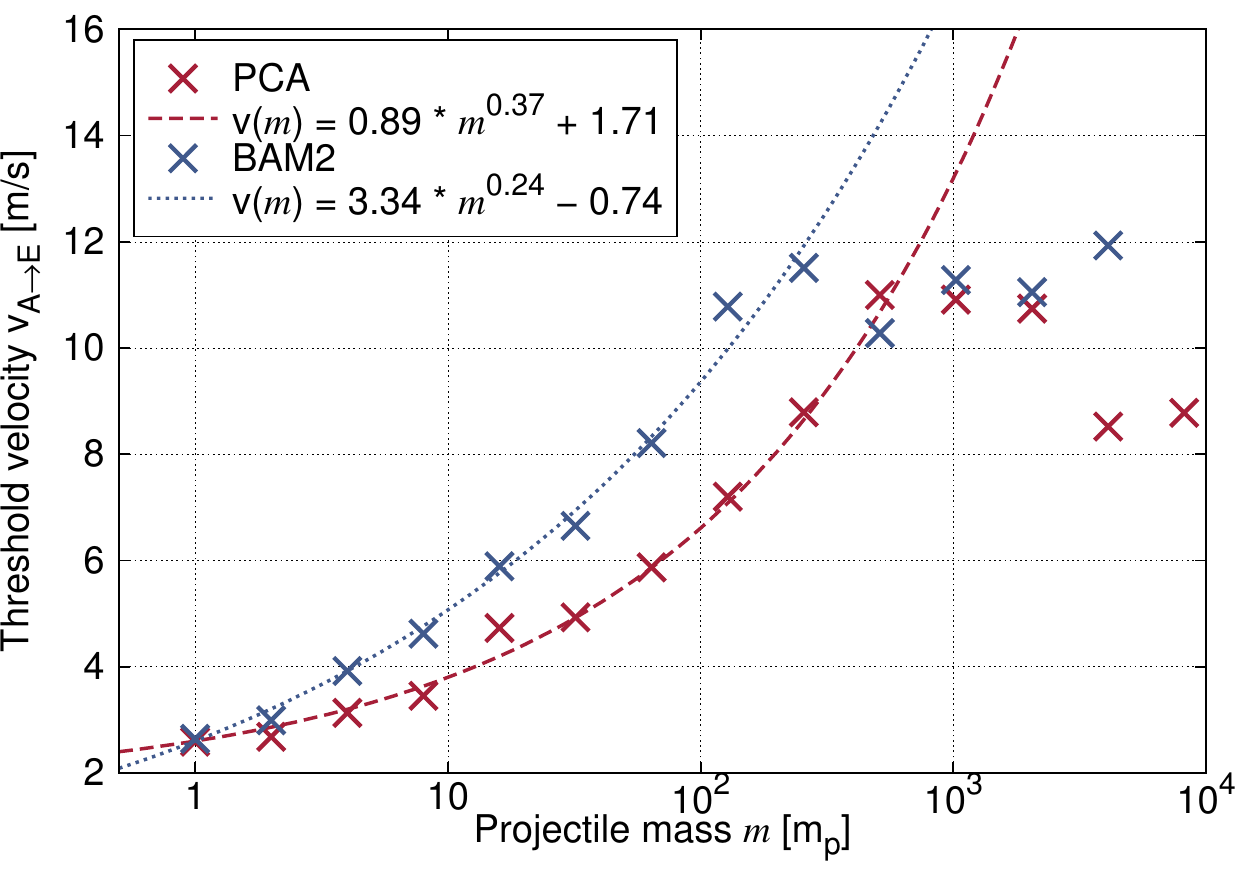}}
\caption{Threshold velocity $v_{\mathrm{A}\rightarrow\mathrm{E}}$ where the transition from accretion to erosion occurs for different projectile sizes (in monomers). The targets are the same PCA and BAM2 aggregates as depicted in Fig.\,\ref{fig:aggregates}. At first, $v_{\mathrm{A}\rightarrow\mathrm{E}}$ increases with growing projectile mass. When the projectiles become too massive $v_{\mathrm{A}\rightarrow\mathrm{E}}$ drops as the collisions enter the fragmentation regime.}
\label{fig:erosion_threshold}
\end{figure}

Independent of the porosity of the target, we find that $v_{\mathrm{A}\rightarrow\mathrm{E}}$ increases significantly with growing projectile mass (see Fig.\,\ref{fig:erosion_threshold}). For projectiles consisting of only a single monomer we find $v_{\mathrm{A}\rightarrow\mathrm{E}} = 2.6\,\mathrm{m s^{-1}}$ and $v_{\mathrm{A}\rightarrow\mathrm{E}} = 2.64\,\mathrm{m s^{-1}}$ for PCA and BAM2 targets, respectively. Indeed, this is equivalent to the critical sticking velocity of two spherical grains predicted using the theory of \citet{krijt2013}. For smaller projectiles between 1 and 256 monomer masses, we determined a fit for the transition velocity
\begin{eqnarray}v_{\mathrm{A}\rightarrow\mathrm{E}, \mathrm{PCA}}(N) = 0.89 N^{0.37} + 1.71 ,\label{eqn:erosion_threshold_fit_PCA}\end{eqnarray}
and
\begin{eqnarray}v_{\mathrm{A}\rightarrow\mathrm{E}, \mathrm{BAM2}}(N) = 3.34 N^{0.24} - 0.74,\label{eqn:erosion_threshold_fit_BAM2}\end{eqnarray}
where $N$ denotes the number of monomers of the projectile (see dashed curve in Fig.\,\ref{fig:erosion_threshold}).

For larger projectiles the measured threshold velocities $v_{\mathrm{A}\rightarrow\mathrm{E}}$ do not follow the fits given in Eqs.\,\ref{eqn:erosion_threshold_fit_PCA} and \ref{eqn:erosion_threshold_fit_BAM2} anymore. This can be explained by the transition from the erosion to the fragmentation regime. In the erosion regime the vast majority of fragments is tiny (below 10 monomers) whereas in the fragmentation regime the impact energy of the projectiles is sufficient to shatter the target into larger fragments (see Fig.\,\ref{fig:erosion_vs_fragmentation}). From the mass $m_\mathrm{T}$ of the target we can estimate the mass ratio, where the collision outcome is dominated by fragmentation. For the data shown in Fig.\,\ref{fig:erosion_threshold}, the critical projectile mass is between $256 - 512\,m_\mathrm{P}$ and $m_\mathrm{PCA} = 8 \cdot 10^4\,m_\mathrm{P}$, $m_\mathrm{BAM2} = 1.5 \cdot 10^5\,m_\mathrm{P}$, where $m_\mathrm{P}$ denotes the mass of a single monomer. Thus, the fragmentation regime is entered when the projectile mass exceeds roughly $0.5\,\%$ of the target mass. 

Though the exact values differ, the evolution of $v_{\mathrm{A}\rightarrow\mathrm{E}}$ for PCA and BAM2 targets is qualitatively very similar (see Fig.\,\ref{fig:erosion_threshold}). Concerning the formation of larger bodies this is a positive result because it indicates that, regardless of the porosity of the target, growth is possible at velocities that are considerably above the sticking velocity of two individual dust grains.

In laboratory experiments, mass growth was found at velocities of about $50\,\mathrm{m s^{-1}}$ \citep{2009MNRAS.393.1584T}. Recently, \citet{Meisner2013} studied high velocity impacts of SiO$_2$ dust aggregates and found that growth is possible at velocities of $\approx 70\,\mathrm{m s^{-1}}$. These velocities are considerably higher than our results. However, the size regime is completely different: The size of the projectiles used by \citet{Meisner2013} is comparable to the size of our target aggregates. Since the maximum value for $v_{\mathrm{A}\rightarrow\mathrm{E}}$ is limited by the onset of fragmentation we expect to observe accretion at higher velocities when using cm- to decimeter-sized target aggregates. Unfortunately, numerical simulations of aggregates of mm size and above are infeasible with the currently available computing power.

\section{Conclusions}
Let us briefly summarize the key results of this work. First of all, in Sect.\,\ref{sec:calibration} we have shown that the JKR description of the repulsion and adhesion between two microscopic silicate grains fails to reproduce the erosion efficiency measured in laboratory experiments by a factor of about 20. By extending the interaction model by a visco-elastic damping force we obtain very good agreement between numerical simulations and laboratory experiments for collision velocities below $30\,\mathrm{m s^{-1}}$. Thus, it is crucial to take this additional damping force into account for any further simulations of dust aggregates in the velocity regime of $\mathrm{m s^{-1}}$.

Secondly, we found that the passivation effect observed in laboratory experiments originates from the artificial setup (see Sect.\,\ref{sec:passivation}). In the context of dust growth in a  protoplanetary disk passivation against erosion does not play an important role. 

In Sect.\,\ref{sec:results} we studied how much different types of aggregates are affected by erosion. Especially the fluffy, fractal aggregates that form during the Brownian motion driven growth phase are prone to erosion. Despite their rather compact surface we find that even the BAM2 aggregates suffer from erosion, though less than the fractal or PCA aggregates. We provide simple recipes to quantify the erosion efficiency for the different aggregate types.

We also examine the influence of the projectile size. Indeed, it turns out that the transition from accretion to erosion is shifted to higher velocities as the projectiles become larger (see Sect.\,\ref{sec:projectile_size}). The possibility of accretion at impact velocities of $20\,\mathrm{m s^{-1}}$ and above helps the growth of larger bodies.

At this point it is hard to judge how these results influence the growth process in protoplanetary disks. For a single impact, we have shown that the erosion efficiency depends on the impact velocity, the structure of the target aggregate as well as the size of the projectile. The prevalence of such impacts is determined by the abundance of small projectiles and the turbulence. As already mentioned in Sect.\,\ref{sec:introduction}, the amount of small grains will be depleted rapidly by collisional growth to larger aggregates \citep{2005A&A...434..971D}. In this work, we have shown that erosion (especially as long as the target aggregates are fluffy and fractal) will produce a steady stream of small dust grains. Therefore it could help to replenish the pool of small grains. The final outcome of this complex interplay of different effects will have to determined by future simulations of the collisional evolution of dust aggregates in protoplanetary disks.

\begin{acknowledgements}
A.\,Seizinger acknowledges the support through the German Research Foundation (DFG) grant KL 650/16. The authors acknowledge support through DFG grant KL 650/7 within the collaborative research group FOR\ 759 {\it The formation of planets}. Dust studies at Leiden Observatory are supported through the Spinoza Premie of the Dutch science agency, NWO.
The authors thank Carsten G\"uttler and J\"urgen Blum (TU Braunschweig) for inspiring discussions.
\end{acknowledgements}

\bibliographystyle{aa}
\bibliography{references}

\end{document}